\documentclass[reprint,superscriptaddres,aps,bbm,pra,longbibliography]{revtex4-1}
\usepackage{graphics}
\usepackage{bm}
\usepackage{lipsum}
\usepackage{epstopdf}
\usepackage{graphicx}
\usepackage{amsmath}
\usepackage{amsthm}
\usepackage{bm}
\usepackage{bbm}
\usepackage{amsmath,amssymb}
\usepackage{physics}
\usepackage{braket}
\usepackage{float}
\usepackage{lipsum}
\newcommand{\ppsi}{\text{\large\ensuremath{\Psi}}}
\begin{document}
\title{History states of one-dimensional Quantum Walks} 
\author{F.\ Lomoc$^1$, A.P.\ Boette$^1$, N. Canosa$^1$, R. Rossignoli$^{1,2}$}
\affiliation{$^1$ Instituto de F\'{\i}sica La Plata, CONICET, and Departamento\ de F\'{\i}sica, Universidad Nacional de La Plata,
C.C. 67, La Plata (1900), Argentina\\
$^{2}$ Comisi\'on de Investigaciones Cient\'{\i}ficas (CIC), La Plata (1900), Argentina}
\begin{abstract} 

We analyze the application of the history state formalism to quantum walks. The formalism allows one to describe the whole walk through a pure quantum history state, which can be derived from a timeless  eigenvalue equation. It  naturally leads to the notion of system-time entanglement of the walk, which can be considered  as a measure 
 of the number of  orthogonal states visited in the walk. We then focus on one-dimensional discrete quantum walks, where it is shown that such entanglement is independent of the initial spin orientation for real Hadamard-type coin operators and real  initial  states (in the standard basis) with definite site-parity.  Moreover, in the  case of an initially localized particle it can be identified with  the  entanglement of the unitary global operator that generates the whole history state,  which is related to  its entangling power and can be analytically evaluated. 
Besides, it is shown that the evolution of the spin subsystem can also be described  through a spin history state with an extended clock. A connection between its average entanglement (over all initial states) and that of the operator generating this state is also derived. 
A quantum circuit for generating the quantum walk history state is  provided as well. 
\end{abstract}
\maketitle

\section{Introduction}
Quantum  Walks (QW) were first introduced in  \cite{YA.93} as a quantum counterpart of classical random walks.  
While related ideas can be traced back to the works of Feynman on the discretized  Dirac equation   \cite{feynman2010quantum}, 
interest on QW has grown enormously in the last decades due to their relevance in the field of quantum computation and information \cite{Ke.03, venegasrev.12}. They are useful for the development of quantum  algorithms 
\cite{Amba.03}, with QW based  methods  \cite{kempe.03,childs.04,nav.8}  achieving similar speedup to that of the renowned Grover search algorithm \cite{GR.97}. It  has also been proven that both continuous  and discrete time QW are universal  for quantum computation \cite{childs.09,ken.10}. They have been employed in network
analysis \cite{nav.9} and quantum simulation \cite{sim.10,sim.11,sim.12,sim.13}, as well as for modeling  some biological  processes \cite{bio.13,bio.14}. 
Besides, QW can be simulated by means of different experimental platforms, such as cold atoms in optical lattices  \cite{expatom.09, expatoms.15, expcold.16},  trapped ions \cite{expTrapionb.09,expTrapionc.10} and  photonic setups   \cite{experim.16,photonic.19,Wang.19,Xu.18,car.20,QQW.20,col.22}.  
 Entanglement in QW is also a topic of interest, so far  mainly focused on that between coin (spin) and position degrees of freedom  \cite{carneiro.05,venegas.05,roma.06,omar.06,maloyer.07,Annabes.10, roma.10, Ide.11, Alles.12, rig.14, Orthey.17,QQW.20,col.22}. 
 
 The aim of this work is to apply the history state formalism \cite{Paw2.83,CR.91,IS.94,Ma.15, Moreva.14, Clean.15,BRGC.16,BR.18} to QW. In this formalism,     originally proposed by Page and Wootters  \cite{Paw2.83}, time is incorporated through a reference quantum clock and the system evolution emerges from an entangled  system-clock history state which fulfills a timeless Wheeler-DeWitt-like equation \cite{dewitt.1967}. The approach has attracted much interest in recent years in different areas, including  nonrelativistic quantum mechanics (QM)  \cite{Ma.15,Moreva.14,Clean.15,BRGC.16,Pa.17,BR.18,Ni.18,Ll.19,DP.19,Cir.22} as well as quantum gravity and relativistic QM   \cite{CR.91,IS.94,Ga.09,zh.18,di.19,Diaz.212}. 

  In this work we will focus on one-dimensional discrete QW 
  \cite{Amba.01,Konno.02,Ke.03,venegasrev.12,tregen.03},  in which a spin $1/2$ quantum particle  undergoes a  unitary evolution in a discrete homogeneous lattice  
  in discrete time steps, according to a translation rule controlled at each step  by the value of its spin component and the action of a quantum gate on the spin.  We will  analyze the walk from the new perspective provided by the history state formalism, here applied in its discrete version  \cite{Clean.15,BRGC.16,BR.18}. The ensuing history state is a pure state of the composite system comprising  both the  position and spin degrees of freedom on one side (system $S$), and  a quantum clock system $T$ on the other. It  contains the information of the whole QW and  satisfies a timeless eigenvalue equation, such that the state of $S$  at step $n$ can be obtained by conditioning the history state with the corresponding clock state. 
  It can be generated from an initial product state through a quantum circuit. 
  
    The corresponding system-time entanglement is a measure of the degree of evolution of $S$, i.e., of the number of orthogonal states visited in the QW, and will be shown to be fully determined by the overlaps between the evolved states. Notice that this entanglement is not the spin-position entanglement usually considered  \cite{carneiro.05,venegas.05,roma.06}.
 We then show that for real  Hadamard-type coin operators such entanglement becomes independent of the initial spin state for a wide class of initial position states, including the standard case of an initially localized particle. In the latter, it is also shown that this 
 entanglement is the operator entanglement of the quantum gate generating the history state from the initial product state. Such entanglement defines in fact the entangling power of this operator, which determines the average over all initial spin states of the history state entanglement. In the case of the quadratic entropy it can be evaluated analytically, and an asymptotic expression for large $N$ can be derived in terms of the coin  operator parameter, showing explicitly  the  deviation from the maximum entropy (i.e., maximally  entangled) limit. The associated entanglement spectrum is also analyzed. 

Besides, it is also possible to define a spin history state by considering a different partition of the whole system, with the spin on  one side and a composite clock (the original clock plus the position degree of freedom) on the other. In the case of an initially localized particle the ensuing spin-rest entanglement 
is shown to be related to that of the  unitary  operator generating  the spin history state. An upper bound to the average spin-rest entanglement is thus obtained, arising from the reduced Schmidt rank of the previous operator. 

  We first briefly review in Section \ref{II} the main features of discrete one dimensional QW and their exact evolution. The  history state formalism for the QW is introduced in section \ref{III}, 
 where the main results, together with analytic expressions for  overlaps and system--time entanglement entropy are provided. 
  The connection with  operator entanglement and the spin history state are discussed in section \ref{IV}.  Illustrative numerical results are shown in both \ref{III} and \ref{IV}.  Finally conclusions are drawn in \ref{V}.

\section{Quantum walks in one dimension\label{II}}
\subsection{Generalities}
Standard one-dimensional quantum  walks  are processes in which a quantum ``particle'' (quantum system) with spin $\frac{1}{2}$ and  hence internal Hilbert space $\mathcal{H}_s = \mathbb{C}^2$, 
moves along a one-dimensional lattice spanned by  position eigenstates $|x\rangle$,  $x\in\mathbb{Z}$, which generate the position Hilbert space $\mathcal{H}_p$. 
The full Hilbert space of the system is then $\mathcal{H}_{S}=\mathcal{H}_p \otimes \mathcal{H}_s$. 
At each time step two operations are performed such that their composition gives the unitary evolution from one time to the next. The first one acts on the spin component (quantum coin) and leaves the position unchanged. This is usually taken as 
a kind of generalized Hadamard transform \cite{tregen.03}, represented in the standard basis (eigenstates $\{|\!\!\uparrow\rangle,|\!\!\downarrow\rangle\}$ of $\sigma_z$)  as 
\begin{equation}C = 
\begin{pmatrix}
\sqrt{\rho} & \sqrt{1-\rho}\, e^{i\gamma} \\
\sqrt{1-\rho}\, e^{i\phi} & -\sqrt{\rho}\,e^{i(\gamma+\phi)}\end{pmatrix}, 
\end{equation}
where $\gamma,\phi$ are arbitrary angles and $0\leq \rho\leq 1$, such that $C^\dag C=\mathbbm{1}$. 
We will here focus on traceless 
real coin  operators, such that it results in a Hadamard-type operator 
\begin{equation} 
C=\begin{pmatrix}
\cos\theta  & \sin\theta  \\
\sin\theta  & -\cos\theta\end{pmatrix}=
\sigma_z\cos\theta +\sigma_x\sin\theta\equiv\sigma_\theta, \label{cm}
\end{equation}
 with $\sigma_x=\begin{pmatrix}
  0 & 1  \\
  1 & 0 \\
 \end{pmatrix}$, $\sigma_y=\begin{pmatrix}
  0 & -\imath  \\
  \imath & 0 \\
 \end{pmatrix}$,  $\sigma_z=\begin{pmatrix}
  1 & 0  \\
  0& -1 \\
 \end{pmatrix}$  the  Pauli matrices and $\cos^2\theta=\rho$. The usual Hadamard gate  is  recovered for $\theta=\pi/4$. 
  Any traceless hermitian unitary coin operator $C$ 
  can be written in the form  \eqref{cm} by adequately chosing the $x$ axis. 
 
The second operation is a conditional one-step displacement to the right (left) if spin is up (down) along $z$, 
generated by the translation operator $T$ ($T^\dag$). 
The total operator generating the step is then given by 
\begin{eqnarray}
U&=&\tfrac{1}{2}\left[T\otimes(\mathbbm{1}+\sigma_z)\sigma_\theta+T^\dag\otimes (\mathbbm{1}-\sigma_z)\sigma_\theta\right]\label{Uy}\\&=&
\sum_{x}[|x+1\rangle\langle x|\otimes |\!\uparrow\rangle\langle\uparrow\!|\sigma_\theta+|x-1\rangle\langle x|\!\downarrow\rangle\langle\downarrow\!|\sigma_\theta]\,,\;\;
\label{Ux}
\end{eqnarray}
and verifies the unitary condition $U^\dag U=\mathbbm{1}_{p}\otimes \mathbbm{1}_{s}$.
\subsection{Exact evolution and overlaps}
Assuming an initial state which is localized in space, i.e., with support in some finite interval, it is convenient, for obtaining a closed exact analytic expression of the evolved states, to consider a finite position basis   $\{|x\rangle,  x=0,\ldots,M-1\}$, 
 with $\langle x|x'\rangle=\delta_{xx'}$, which contains the initial state as well as the evolved states up to, say, $N$ steps (i.e.\ $M>2N$ for an initially localized particle, where $N$ is the number of steps). 
 Then we define the associated discrete Fourier transformed (DFT) basis   
 \begin{equation}
     |k\rangle=\frac{1}{\sqrt{M}}\sum_{x=0}^{M-1}e^{i2\pi xk/M}|x\rangle,\;\;k=0,\ldots,M-1\,,
 \end{equation}
satisfying $\langle k|k'\rangle=\delta_{kk'}$ and $|\!-k\rangle=|M\!-\!k\rangle$. These states are the eigenstates of the cyclic  translation operator defined by $T|x\rangle=|x+1\rangle$, with 
$T|M-1\rangle=|0\rangle$, such that  
\begin{equation}
    T|k\rangle=e^{-i2\pi k/M}|k\rangle\,.
\end{equation}
The operator \eqref{Ux} can then be rewritten as 
\begin{subequations}
\label{5}\begin{eqnarray}
    U&=&\exp[-i \frac{2\pi}{M} K\otimes \sigma_z] (\mathbbm{1}\otimes \sigma_\theta)\\&=&
\begin{pmatrix}e^{-i2\pi K/M}&0\\0&e^{i2\pi K/M}\end{pmatrix}   \begin{pmatrix}\cos\theta&\sin\theta\\\sin\theta&-\cos\theta\end{pmatrix}\label{Uib}\,,
\end{eqnarray}
\end{subequations}
where $K$ is the discrete ``momentum'' operator satisfying  $K|k\rangle=k|k\rangle$, such that $T=\exp[-i2\pi K/M]$.

For each value $k$ of $K$, \eqref{Uib} 
represents, in the standard spin basis, a unitary operator $U_k$ in spin space fulfilling ${\rm Det}[U_k]=-1$, with eigenvalues 
\begin{subequations} \label{8}
\begin{eqnarray}
\lambda_k^{\pm}&=&\pm e^{\mp i \omega_k}\label{8a}\\
&=&\pm\sqrt{1-\cos^2\theta\sin^2\phi_k}-i\cos\theta\sin\phi_k\,,
\label{8b}
\end{eqnarray}
\end{subequations}
where $\phi_k=2\pi\,k/M$, and eigenvectors 
\begin{eqnarray}
    |s^{\pm}_k\rangle&=&\alpha^\pm_k|\!\!\uparrow\rangle+\beta^\pm_k|\!\!\downarrow\rangle\,,\;\;\frac{\beta^\pm_k}{\alpha^\pm_k}=
\frac{e^{i\phi_k}\lambda_k^\pm-\cos\theta}{\sin\theta}\,,\;\;\;\;\label{skst}
\end{eqnarray} 
satisfying $\langle s_k^{\nu}|s_{k'}^{\nu'}\rangle=\delta_{kk'}\delta^{\nu\nu'}$. 
 Thus, 
 \begin{subequations}\label{10}
 \begin{eqnarray}U&=&\sum_{k=0}^{M-1} |k\rangle\langle k|\otimes U_k\,,\label{Ukk}\\
 U_k&=&e^{-i\omega_k}|s_k^+\rangle\langle  s_k^+|-e^{i\omega_k}|s_k^-\rangle\langle  s_k^-|\,.\label{10b}
 \end{eqnarray}
 \end{subequations}
Whereas in the position representation \eqref{Ux} it is the spin which appears as controlling the position displacement, in the momentum representation \eqref{Ukk}  based on the eigenbasis of $T$,  it is the momentum $K$ which controls the spin evolution at each step. 

Using \eqref{10} we can now determine the evolution of a general initial product state 
\begin{subequations}
\begin{align}|\Psi_0\rangle&=|\psi_0\rangle\otimes |\chi_0\rangle\,,\\
|\psi_0\rangle&=\sum_x \psi_0(x)|x\rangle=\sum_k c_k |k\rangle\,,\label{13b}\\
|\chi_0\rangle&=\alpha|\!\!\uparrow\rangle+\beta|\!\!\downarrow\rangle\,,
\label{13c}\end{align}\label{13}
\end{subequations}
$\!$where $\psi_0(x)$ is the initial position state, with  
\begin{equation}
c_k=\langle k|\psi_0\rangle=\frac{1}{\sqrt{M}}\sum_{x}e^{-i2\pi x k/M}\psi_0(x)\label{DFTk}\end{equation} the DFT of $\psi_0(x)$ (sums over $k,x$ are from $0$ to $M-1$)  and $\ket{\chi_0}$
the initial spin state. The state after $n$ steps is 
\begin{align}
    |\Psi_n\rangle=U^n|\Psi_0\rangle
    =\sum_k c_k|k\rangle\otimes 
    U^n_k|\chi_0\rangle\label{Un}\,,
     \end{align} 
where the evolved spin state for momentum $k$ is 
    \begin{align}
U_k^n|\chi_0\rangle&= 
e^{-in\omega_k}\langle s_k^+|\chi_0\rangle|s_k^+\rangle+e^{in\omega_k}\langle s_k^-|\chi_0\rangle|s_k^-\rangle\,.
\end{align}
 
 A quantity of most importance in this work is the overlap between the  evolved states, \begin{subequations}\label{ov}\begin{eqnarray}\langle\Psi_{n'}|\Psi_n\rangle&=&\langle \Psi_0|U^{n-n'}|\Psi_0\rangle=
 \langle\Psi_0|\Psi_{n-n'}\rangle\label{ov1}\\
&=&\sum_k |c_k|^2\langle \chi_0|U_k^{n-n'}|\chi_0\rangle\label{ov2}\\
&=&\sum_k |c_k|^2\!\!\sum_{\nu=\pm 1}
\!\!\nu^{n-n'} e^{i\nu (n'-n)\omega_k}|\langle s_k^\nu|\chi_0\rangle|^2\,,
\;\;\;\;\;\;\;\;\label{ov3}\end{eqnarray}
\end{subequations}
which depends just on $n-n'$. 
 
For future use we notice that  for $-k\equiv M-k$, 
     $\lambda_{-k}^{\pm}=\lambda_k^{\pm\,*}$ ($\omega_{-k}=-\omega_k$)    and  $|s^\pm_{-k}\rangle=|s^\pm_{k}\rangle^*$ in the standard basis ($\alpha_{-k}^\pm=\alpha_k^{\pm\,^*}$, $\beta_{-k}^\pm=\beta_k^{\pm\,*}$ in \eqref{skst}). 
     Moreover,  
\begin{equation}
U_{-k}=-\sigma_y U_k\sigma_y\label{sgmy}\,,
\end{equation}
entailing $U_{-k}^n=(-1)^n \sigma_yU_{k}^n\sigma_y$. It is also evident  
from \eqref{Uib} that for $M$ even, 
\begin{equation}
U_{k+M/2}=-U_k\,.\label{sm2}
\end{equation}

We finally notice that in the special case  $\theta=\pi/2$, $\lambda_k^{\pm}=\pm 1$ ($\omega_k=0$) and 
$U_k^2=\mathbbm{1}_s$ $\forall$ $k$, i.e.\  $U^2=\mathbbm{1}$: 
the system always returns to its initial configuration after two steps, 
 as  $C=\sigma_x$ (Eq.\ \eqref{cm}) flips  the coin at each step.

\section{Quantum walk history states\label{III}}
\subsection{Definition and main properties}
Let us now consider a quantum walk with $N-1$ steps, starting from an initial state $|\Psi_0\rangle$ at time $t_0=0$  and ending in a  state $|\Psi_{N-1}\rangle$ at time $t_{N-1}=(N-1)\tau$, where $\tau$ is a certain time scale. We then consider a quantum clock system $T$ with an orthogonal set of states  
$\ket{n}$, $n=0,\ldots,N-1$, 
representing the scaled time $n=t_{n}/\tau$ at which  the $n^{\rm th}$ step takes place. They are eigenstates of a clock operator $T_c$ satisfying $T_c|n\rangle=n|n\rangle$, and  could represent, for instance,  the scaled position of the clock's needle.  

We now define the quantum walk history state as  
\begin{equation}
\left|\ppsi\right\rangle=\frac{1}{\sqrt{N}}\sum_{n=0}^{N-1}|\Psi_n\rangle\otimes \ket{n}\,,\label{psih}
\end{equation}
where $|\Psi_n\rangle$ is the system state  \eqref{Un} at step $n$.  The state \eqref{psih} contains the whole information of the walk. For example, the  time average of an observable $O_{S}$ of the particle   over the complete walk can be expressed as 
\begin{subequations}
\label{av}\begin{eqnarray}\langle O_{S}\rangle&:=&\frac{1}{N}\sum_{n=0}^{N-1}\langle \Psi_n|O_S|\Psi_n\rangle\\
&=&\langle\ppsi|O_{S}\otimes\mathbbm{1}_T|\ppsi\rangle\label{Om}\,,\end{eqnarray}
\end{subequations} 
whereas matrix elements between system states at any two times  can be written as 
\begin{align}\langle\Psi_{n'}| O_{S}|\Psi_n\rangle&=N\langle\ppsi|(O_S\otimes |n'\rangle\langle n|)|\ppsi\rangle\,.\end{align}

\begin{figure}[h]
    \centering
    \includegraphics[width=7.5cm]{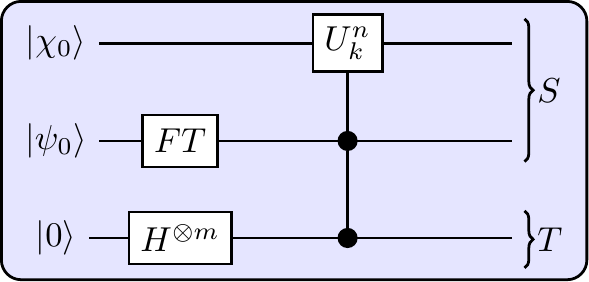}
    \caption{Schematic circuit representing the generation of the history state \eqref{psih} through Eq.\ \eqref{Un}: The Quantum Fourier Transform (FT) is applied to the initial position state $|\psi_0\rangle$, while $H^{\otimes m}$ denotes the
Hadamard operator over $m$ qubits with $2^m = N$, such that $H^{\otimes m}|0\rangle=\frac{1}{\sqrt{N}}\sum_{n=0}^{N-1}|n\rangle$. Finally the controlled-$U$ gate $U_k^n$ acts over the initial spin state $\ket{\chi_0}$.}
    \label{fig1}
\end{figure}

Using Eqs.\ \eqref{Un} and \eqref{psih}, the quantum walk history state can be generated from an initial product  state $|\psi_0\rangle|\chi_0\rangle|n=0\rangle$ through the schematic circuit of Fig.\ \ref{fig1}, where $FT$ implements the  transformation $|x\rangle\rightarrow\frac{1}{\sqrt{M}}\sum_ke^{-i2\pi xk/M}|k\rangle$.  For a particle  initially localized at $x=0$, $c_k=1/\sqrt{M}$  $\forall$ $k$ and one may replace the FT gate by a simpler Hadamard gate  $H^{\otimes m'}$, with $2^{m'}=M$.  

History states actually {\it determine}  the system evolution when they satisfy a proper {\it timeless} eigenvalue equation \cite{BR.18}. Defining the system-clock unitary  operator 
\begin{equation}
    {\cal U}=\sum_{n=1}^{N}U_{n,n-1}\otimes |n\rangle\langle n-1|, \label{Ung}
    \end{equation}
where $U_{n,n-1}=U$ for $1\leq n\leq N-1$ 
and (identifying $|N\rangle$ with  $|0\rangle$) 
$U_{N,N-1}=(U^{N-1})^\dag$,  
the state \eqref{psih} is first seen to be an exact eigenstate of  ${\cal U}$ with eigenvalue $1$: 
\begin{equation}
    {\cal U}|\ppsi\rangle =|\ppsi\rangle\,.\label{Up}
    \end{equation}

 Conversely, if $|\ppsi\rangle$ is {\it any} state satisfying Eq.\ \eqref{Up}, i.e.\ a state which remains fully invariant under ${\cal U}$,  then  
 \begin{equation}\langle n|\ppsi\rangle=\langle n|{\cal U}|\ppsi\rangle=U_{n,n-1}\langle n-1|\ppsi\rangle\,,\end{equation}
 implying that system states $|\Psi_n\rangle:=\sqrt{N}\langle n|\ppsi\rangle$ will fulfill the discrete unitary evolution  $|\Psi_n\rangle=U_{n,n-1}|\Psi_{n-1}\rangle$, i.e.,  Eq.\ \eqref{Un} for constant $U_{n,n-1}=U$ ($1\leq n\leq N-1$). Normalization $1=\langle\ppsi|\ppsi\rangle=\sum_n \langle\ppsi|n\rangle\langle n|\ppsi\rangle$ then entails $\langle\Psi_n |\Psi_n\rangle=1$. The whole system evolution up to step $N-1$ is thus  determined by Eq.\ 
\eqref{Up} and the initial state $|\Psi_0\rangle$. The same holds for  general unitaries  $U_{n,n-1}$ provided   $U_{N,N-1}\ldots U_{1,0}=\mathbbm{1}_S$. 

Writing ${\cal U}=\exp[-i{\cal J}]$, Eq.\ \eqref{Up} implies ${\cal J}|\Psi\rangle=0$ (or $2k\pi$, $k$ integer),  which is a discrete  Wheeler-DeWitt-type equation \cite{dewitt.1967}. In fact, if 
 $U_{n,n-1}=U$ $\forall$ $n=1,\ldots,N$ (such that $U^N=\mathbbm{1}$) then ${\cal J}=H\otimes\mathbbm{1}_{T}+\mathbbm{1}_S\otimes P_c$, 
 where $e^{-iH}=U$ determines the step evolution in $S$  and $e^{-iP_c}=\sum_{n=1}^N|n\rangle\langle n-1|$  generates the unit translation in the time basis ($|n-1\rangle\rightarrow |n\rangle$)  
 \cite{BR.18}. 
 
 We also mention that the other eigenvalues of ${\cal U}$ are $e^{-i2\pi k/N}$ ($k=0,\ldots,N-1$) \cite{BRGC.16,BR.18}. Hence, 
 $|\Psi\rangle$ can also be seen as the ground state of the hermitian operator $-\frac{1}{2}({\cal U}+{\cal U^\dag})$, which has  eigenvalues $-\cos \frac{2\pi k}{N}$. This enables  the use  of variational methods for its determination 
 \cite{Clean.15,Cir.22}.
 
\subsection{System-time entanglement}
The entanglement $E(S,T)$ of the history state \eqref{psih} 
 can be  regarded as a measure of the {\it ``degree of evolution''} of  system $S$, i.e., of the number of orthogonal states visited in the walk:  If $|\Psi_0\rangle$ is an eigenstate of $U$, i.e.\ a {\it stationary}  state satisfying  $U|\Psi_0\rangle=e^{i\gamma}|\Psi_0\rangle$, the history state becomes {\it separable}:    $|\ppsi\rangle=|\Psi_0\rangle(\frac{1}{\sqrt{N}}\sum_n e^{in \gamma}|n\rangle)$,  and $E(S,T)=0$ \cite{BRGC.16}.  This case arises here in finite  cyclic realizations when    $|\psi_0\rangle$ in (14) remains invariant under translations, i.e., when it is a state $|k\rangle$, and  $|\chi_0\rangle$ coincides with one of the eigenstates $|s_k^{\pm}\rangle$ of $U_k$, such that  $U^n|\Psi_0\rangle=e^{\mp in\omega_k}|\Psi_0\rangle$ for  $|\Psi_0\rangle=|k\rangle\otimes|s_k^\pm\rangle$. 

The opposite situation  is an evolution where at each  step the system evolves into a new  orthogonal state  
such that $\langle \Psi_n|\Psi_{n'}\rangle=\delta_{nn'}$, in which case $E(S,T)$ is {\it maximum}. This evolution arises here in the trivial limit  $\theta\rightarrow 0$ for an initially localized state $\psi_0(x)=\delta_{xx_0}$ with definite spin along the $z$ axis ($\sigma_z|\chi_0\rangle=\pm|\chi_0\rangle$),  such that the particle advances always in the same direction.

In the general case 
$E(S,T)$ will be  determined by the overlaps \eqref{ov}. 
We can expand \eqref{psih}  as ($\nu=\pm 1$) 
\begin{align}
|\ppsi\rangle&=\frac{1}{\sqrt{N}}\sum_{k,\nu,n}\langle k,s_k^\nu|\Psi_n\rangle |k,s_k^\nu\rangle\otimes |n\rangle\nonumber\\&=
\sum_{m=1}^{n_s} \sqrt{\lambda_m}|m_S\rangle\otimes |m_T\rangle\,,\label{SD}
\end{align}
where $|k,s_k^\pm\rangle=|k\rangle\otimes |s_k^\pm\rangle$ and \eqref{SD} is its {\it Schmidt representation}, obtained from the singular value decomposition (SVD) $M=VDW^\dag$ of the matrix  of elements  
$M_{k\nu,n}=\langle k,s_k^\nu|\Psi_n\rangle/\sqrt{N}$. 
Here $D_{mm'}=\sqrt{\lambda_m}\delta_{mm'}$ with $\lambda_m$ the eigenvalues of the positive semidefinite matrices $MM^\dag$ or equivalently $M^\dag M$ (fulfilling ${\rm Tr}\,M^\dag M=\sum_m \lambda_m=1$) while $V$, $W$ are unitary matrices diagonalizing $MM^\dag$ and $M^\dag M$ respectively, satisfying $VD=MW$. 

In \eqref{SD}  $|m_{S}\rangle=\sum_{k,\nu}V_{k\nu,m}|k,s_k^\nu\rangle$, 
$|m_T\rangle=\sum_n W^{*}_{nm}|n\rangle$ are orthogonal system and  clock states respectively ($\langle m_{S(T)}|m'_{S(T)}\rangle=\delta_{mm'}$) while  $n_s$ is the {\it Schmidt rank}, i.e.\   the number of nonzero eigenvalues $\lambda_m$, which is just the rank of the matrix $M$. $|\ppsi\rangle$ is then entangled iff $n_s\geq 2$ and separable (product state) if $n_s=1$. 

Eq.\ \eqref{SD} shows that the states $|m_{S(T)}\rangle$ are the eigenvectors of the reduced system and clock states  
\begin{equation}\rho_{S(T)}={\rm Tr}_{T(S)}\,|\ppsi\rangle\langle \ppsi|=\sum_{m=1}^{n_s}p_m|m_{S(T)}\rangle\langle m_{S(T)}|\,,\label{rhoST}\end{equation} 
which determine the average along the walk  of any local observable ($\langle \ppsi|O_S\otimes\mathbbm{1}|\ppsi\rangle\!\!=\!\!{\rm Tr}_S\,\rho_S O_S$), and have the same nonzero eigenvalues $\lambda_m$.   Their entropies  
are then identical and define 
the {\it entanglement entropy} of the history state $|\ppsi\rangle$ (system-time entanglement \cite{BRGC.16})
\begin{equation}E(S,T)=S(\rho_S)=S(\rho_T)=\sum_{m}f(\lambda_m)\,,\label{EST}\end{equation}
where the last expression holds for a general trace-form entropy 
$S(\rho)={\rm Tr}\,f(\rho)$, where  $f$ is a concave nonnegative  function satisfying $f(0)=f(1)=0$.  Eq.\ \eqref{EST} 
vanishes iff the history state is  separable. 

The reduced states \eqref{rhoST} can be here also  written  as
\begin{subequations}\begin{eqnarray}
\rho_{S}&=&
\frac{1}{N}\sum_{n=0}^{N-1}|\Psi_n\rangle\langle\Psi_n|\,,
\label{rhoS}\\
    \rho_{T}&=&
    \frac{1}{N}\sum_{n,n'}\langle \Psi_{n'}|\Psi_n\rangle|n\rangle\langle n'|\,.\label{rhoT}
\end{eqnarray}\label{rhor}
\end{subequations}
Eq.\  \eqref{rhoT} shows explicitly that $\rho_T$, and hence its eigenvalues $\lambda_m$ and the entanglement \eqref{EST} of the history state,  {\it are fully determined by the  overlaps
\eqref{ov}}.  

The standard choice for $S$ is the von Neumann entropy  
\begin{equation}
S(\rho)=-{\rm Tr}\,\rho\log_2\rho\label{VNS},
\end{equation}
which will satisfy \begin{equation}0\leq E(S,T)\leq \log_2 N\label{Svb},\end{equation} 
such that 
$2^{E(S,T)}$ is essentially a measure of the number of distinct orthogonal states visited in the evolution.  Another convenient choice  is the quadratic entropy (also denoted as  linear entropy or $q=2$ Tsallis entropy \cite{TS.09}), which is simply determined by the purity ${\rm Tr}\,\rho^2$: 
 \begin{equation} 
 S_2(\rho)=1-{\rm Tr}\,\rho^2\label{S21}\,,\end{equation}
 and corresponds to $f(\rho)=\rho(\mathbbm{1}-\rho)$. It does not require the explicit determination  of eigenvalues and can be measured without requiring a full state tomography \cite{Naka.12}. It can  be here directly evaluated in terms of the overlaps \eqref{ov}: Using \eqref{rhor} 
 together with \eqref{EST} and \eqref{S21} we obtain 
 \begin{subequations}
 \label{S22}
 \begin{eqnarray}
 E_2(S,T)&=&1-\frac{1}{N^2}\sum_{n,n'}|\langle \Psi_{n'}|\Psi_n\rangle|^2\\
 &=&1-\frac{1}{N}[1+2\sum_{n=1}^{N-1}(1-\frac{n}{N})|\langle\Psi_0|\Psi_n\rangle|^2]\,,\label{S222}\;\;\;\;\;\;\;\end{eqnarray}
 \end{subequations}
 where \eqref{S222} holds when  $|\langle \Psi_{n'}|\Psi_{n}\rangle|$ depends just on  $|n-n'|$, as  in Eq.\ \eqref{ov} 
 (with factor $\propto 1-\frac{n}{N}$ accounting for the pertinent multiplicity). 
It obviously satisfies \begin{equation}0\leq E_2(S,T)\leq 1-\frac{1}{N},\label{S2b}\end{equation} 
such that $\frac{1}{1-E_2(S,T)}$ is here the effective number of orthogonal states visited. A 
directly 
related quantity is the $q=2$ Renyi entropy  
$S_2^R(\rho)=-\log_2(1-S_2)=-\log_2 {\rm Tr}\,\rho^2$ \cite{Beck.93}, which satisfies the same bound \eqref{Svb}.
 
The upper limit in \eqref{Svb}--\eqref{S2b} is reached for $\theta\rightarrow 0$ in \eqref{cm} and an initially localized particle with definite $\sigma_z$.  On the other hand, for $\theta=\pi/2$ the evolution becomes periodic with period $2$ (as $U^2=\mathbbm{1}$) and hence 
the system-time entanglement entropy will stay  trivially bounded $\forall$ $N$:  
\begin{equation}
E(S,T)\leq 1\,,\;\;\;E_2(S,T)\leq 1/2\;\;\;(\theta=\pi/2), 
\end{equation}
with the upper limit reached for an orthogonal intermediate state. Thus,  by varying $\theta$ in the interval $[0,\pi/2]$ we can reach, 
for an initially localized particle, all possible  rates  of system-time entanglement increase with $N$, from the  maximum rate for $\theta=0$ to the null increase  for $\theta=\pi/2$,   entailing in general a decrease of $E(S,T)$ with increasing $\theta$ in this interval.   
 
\subsection{Independence of system-time entanglement from initial spin state} 
\subsubsection{Real initial  states with definite site  parity}
We now examine the entanglement \eqref{EST} for some general types of initial states $|\Psi_0\rangle$.   We first notice that  if $|\Psi_0\rangle$ has a {\it definite position parity},  such that its support are just even (or odd) sites $x$,  
  \begin{equation}(e^{i\pi X}\otimes\mathbbm{1})|\Psi_0\rangle=\pm|\Psi_0\rangle,\label{P0}\end{equation}
  where $X$ is the discrete position operator $X|x\rangle=x|x\rangle$, the overlap $\langle\Psi_n|\Psi_{n'}\rangle$ will vanish for $n-n'$ odd since at each step the particle will move to sites of opposite parity: 
  \begin{equation} (e^{i\pi X}\otimes \mathbbm{1})|\Psi_n\rangle=\pm (-1)^n|\Psi_n\rangle\,.\label{Pn}\end{equation}
     Eq.\ \eqref{P0} is  trivially fulfilled  for an initially localized particle $\psi_0(x)=\delta_{x,x_0}$. 
 In momentum space, Eq.\  \eqref{P0} implies $c_{k+M/2}=\pm c_k$ ($M$ even) and 
\eqref{Pn} follows from  \eqref{sm2} and  \eqref{Un}. 

      Then,    for  {\it real} $\psi_0(x)$  and $\alpha,\beta$   in \eqref{13}, such that $c_{-k}=c_k^*$ and 
   $\langle s_{-k}^\pm|\chi_0\rangle=\langle s_{k}^\pm|\chi_0\rangle^*$,
   we obtain, using $\sum_{\nu=\pm}|\langle s_k^\nu|\chi_0\rangle|^2=1$ and Eq.\ \eqref{ov3},
\begin{eqnarray}\!\!\langle \Psi_{0}|\Psi_n\rangle&=&
\sum_{k}|c_k|^2\cos(n\omega_k)(|\langle s^+_k|\chi_0\rangle|^2+(-1)^n|\langle s^-_k|\chi_0\rangle|^2)\nonumber\\
&=&\left\{\begin{array}{ccl}
\!\!\sum\limits_k|c_k|^2\cos(n\omega_k)\!\!&,&\!\!n\,{\rm even}\\0\!&,&\!\!n\,{\rm odd}\end{array}\right.\!.\label{crw}
\end{eqnarray}
Thus, the overlap  becomes {\it independent of the initial (real) spin state} $|\chi_0\rangle$ $\forall$ $n$. This implies  a system-time entanglement entropy {\it independent} of $|\chi_0\rangle$.   

Previous result can be seen more clearly using  Eq.\ \eqref{sgmy} and \eqref{ov2}: since  $\sigma_y|\chi_0\rangle$ is orthogonal to $|\chi_0\rangle$ 
for $|\chi_0\rangle$ real  ($\alpha,\beta$ real in \eqref{13c},  equivalent to  $\langle\bm{\sigma}\rangle$ in the $x,z$ plane) and $\langle\chi_0|U_{-k}^n|\chi_0\rangle=(-1)^n\langle{\chi}_0|\sigma_y U_k^n\sigma_y|{\chi}_0\rangle$, for $|c_{-k}|=|c_k|$ and $n$ even we obtain  
\begin{eqnarray}
\langle\Psi_{0}|\Psi_{n}\rangle&=&
\frac{1}{2}\sum_{k}|c_k|^2
(\langle \chi_0|U_k^n|\chi_0\rangle+\langle{\chi}_0|\sigma_yU_k^n\sigma_y|\chi_0\rangle)\nonumber\\
&=&\frac{1}{2}\,\sum_k |c_k|^2\,{\rm Tr}\,U_k^{n}\label{35}
={\textstyle\frac{1}{2}}\langle \psi_0|{\rm Tr}_{s}\,U^{n}|\psi_0\rangle\,,
\;\;\;\;\;\end{eqnarray}
which shows that the even overlap depends just on the {\it trace} of $U_k^{n}$, i.e.\ on the partial trace of $U^n$ over spin,  and is hence  independent of the initial spin state $|\chi_0\rangle$. Eq.\ \eqref{35} leads again to \eqref{crw} for $n$ even and {\it also} $n$ odd,  as ${\rm Tr}\,U_{k}^n =e^{-in\omega_k}+(-1)^n e^{in\omega_k}=(-1)^n{\rm Tr}\,U^n_{-k}$ and hence the sum in \eqref{35} vanishes for $n$ odd when $|c_{-k}|=|c_k|$. 

With previous expressions, the sum over $N$ in the quadratic system-time entanglement \eqref{S222} can be evaluated analytically:  
\begin{subequations}\label{S2N}\begin{eqnarray}\!E_2(S,T)
&=&{1-\!\!\!\sum\limits_{\genfrac{}{}{0pt}{}{k,k'}{\nu=\pm 1}}\!
\!\frac{|c_kc_{k'}|^2\sin^2\frac{N(\omega_k+\nu\omega_{k'})}{2}}{N^2\sin^2(\omega_k+\nu\omega_{k'})}},\;N\,{\rm even}\;\;\label{even}\\&=&{1-\!\!\!\!\!\!\!\sum\limits_{\genfrac{}{}{0pt}{}{k,k'}{\nu,\nu'=\pm 1}}\!\!\!\!
\!\frac{|c_kc_{k'}|^2\sin^2\frac{(N+\nu')(\omega_k+\nu\omega_{k'})}{2}}{2N^2\sin^2(\omega_k+\nu\omega_{k'})}},\;N\,{\rm odd}\;\;\;\;\;\;\;\;\;\label{odd}
\end{eqnarray}
\end{subequations}
Here $\frac{\sin^2 mu}{m^2\sin^2 u}$  is understood as its limit $1$ if $\sin u=0$ ($m$ integer),   being a  
polynomial of degree $m-1$  in $\cos u$ ($\frac{\sin (n+1)u}{\sin u}=
\sum_{k=0}^{\lfloor{n/2}\rfloor}(-1)^k\binom{n-k}{k}(2\cos u)^{n-2k}$ is a Chebyshev polynomial of the second kind \cite{Abra.65}).
  
As a check,  for $\theta=\pi/2$, $\omega_k=0$ $\forall$ $k$ and hence, for any  initial $\psi_0(x)$ with definite parity, \eqref{even}--\eqref{odd} lead to 
\begin{eqnarray}
E_2(S,T)&=&\left\{\begin{array}{ccl}\frac{1}{2}&,&N\,{\rm even}\\
\frac{1}{2}(1-\frac{1}{N^2})&,&N\,{\rm odd}\\\end{array}\right.\;\;\;(\theta=\pi/2)\,.\;\;\label{E2th2}
\end{eqnarray}
This means that the system just moves between  two orthonormal states (periodic evolution) as previously stated,  entailing a non-increasing system-time entanglement entropy. 
 The spectrum of $\rho_S(T)$ is simply $(\frac{1}{2},\frac{1}{2})$ for $N$ even and $(\frac{N+1}{2N},\frac{N-1}{2N})$ for $N$ odd.

\subsubsection{Initially localized particle}
For  $\psi_0(x)=\delta_{xx_0}$,   $c_k=\frac{1}{\sqrt{M}}e^{-i2\pi x_0k/M}$ 
and $|c_k|^2=\frac{1}{M}$ $\forall\,k$. Overlaps and  system-time entanglement are then determined  just by the full trace of $U^n$ and hence the coin operator angle $\theta$, as implied by \eqref{crw}-\eqref{35}: 
\begin{equation}\langle \Psi_{0}|\Psi_n\rangle
=\frac{{\rm Tr}\,[U^{n}]}{2M}=\left\{\begin{array}{cl}\!\!\frac{1}{M}\!\sum_k
\cos(n\omega_k),
&n\,{\rm even}\\0,&n\,{\rm odd}\end{array}\right..\label{cw}\end{equation}

 \begin{figure}[h]
    \centering
    \includegraphics[width=.5\textwidth]{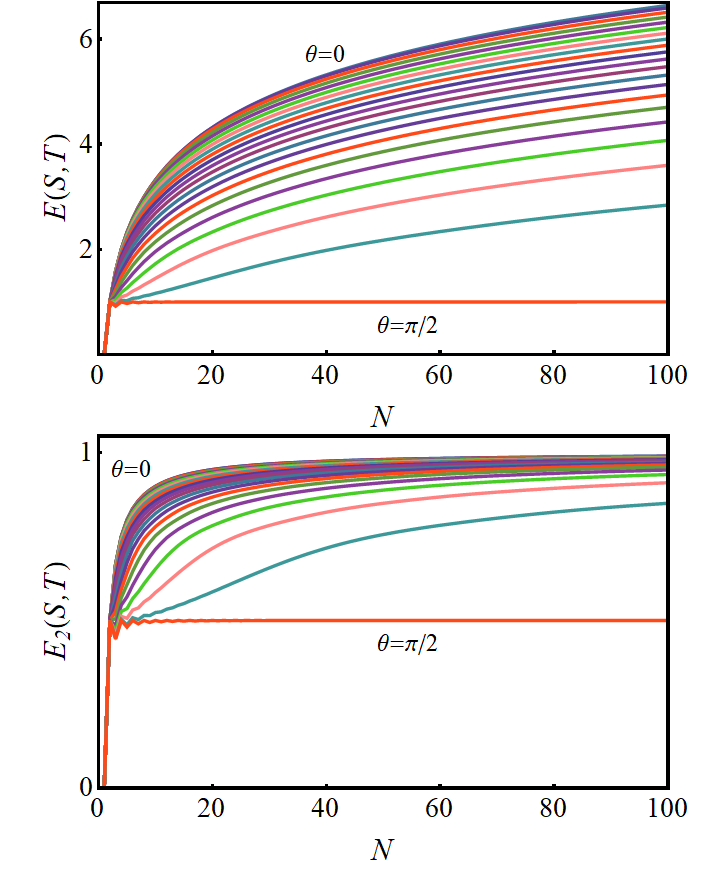}
    \caption{System-time entanglement according to the von Neumann (top) and  quadratic (bottom) entropies, as a function of the total number $N$ of steps of the quantum walk for different values $\theta=0,\pi/20.,\ldots,\pi/2$, of the coin operator parameter,  for an initially localized particle and any real initial  spin state.}
    \label{fig2}
\end{figure}

\begin{figure}
    \centering
    \includegraphics[width=.5\textwidth]{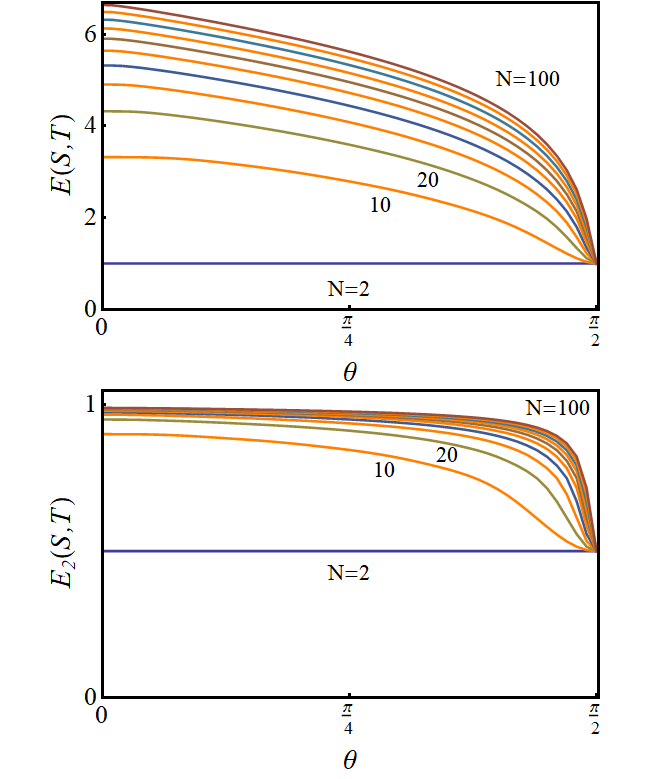}
    \caption{System-time entanglement according to the von Neumann (top) and quadratic (bottom) entropies,  as a function of the coin operator angle $\theta$ for different number of steps $N=2,10,20,\ldots,100$,  for the same initial states of Fig.\ \ref{fig2}.}     \label{fig3}
\end{figure}

We can now evaluate \eqref{cw} {\it exactly $\forall$ $\theta$ and $N$}, 
using \eqref{8}. 
Since the result is independent of $M$ (for $M>2N$ if  $x_0=M/2$) it can be obtained either summing over $k$  or letting  $M\rightarrow\infty$ and replacing the  sum  by an integral over $\phi=2\pi k/M$ (with $d\phi=2\pi/M)$.  We obtain \begin{subequations}
\label{pl}
\begin{eqnarray}
\langle\Psi_{0}|\Psi_{2n}\rangle
&=& P_{n}(\cos^2\theta)\,,\label{pl1}\\
P_n(u)&=& {\textstyle\sum\limits_{j=0}^n(-1)^j \binom{n}{j}\binom{n+j-1}{j}}u^j
\label{pl11}\\
&=&1-n^2u+\tfrac{1}{4}n^2(n^2-1)
u^2+\ldots\hspace*{.5cm}\label{pl2}
\end{eqnarray}
\end{subequations}
where $P_n(u)=F_{2,1}(-n,n,1,u)=P^{0,-1}_n(1-2u)$ 
is a polynomial of degree $n$ in $u$,  with $F_{2,1}$ the  hypergeometric function and 
$P^{\alpha,\beta}_n$ the Jacobi polynomial \cite{Abra.65}. 
 It satisfies $P_n(0)=1$ and $P_n(1)=0$ $\forall$ $n\geq 1$. 

For {\it large} $n$ and $\theta\in(0,\pi/2)$,   we can also  obtain from  \eqref{pl} and the asymptotics of Jacobi polynomials, the {\it exact  asymptotic expression} 
\begin{equation}
\langle\Psi_0|\Psi_{2n}\rangle\approx 
(-1)^n\sqrt{\frac{\tan\theta}{n\pi}}\cos(2n\theta+\pi/4)+O(n^{-3/2})
\label{N}\,,\end{equation}
which shows that the overlap  fades away  as $n^{-1/2}$ for large $n$ and its modulus essentially increases as $\tan^{1/2}\theta$ for increasing  $\theta\in(0,\pi/2)$.

Through  Eqs.\ \eqref{cw}--\eqref{pl} the quadratic system-time entanglement entropy \eqref{S22} can be evaluated exactly as: 
\begin{subequations}
\label{E2N}
\begin{eqnarray}
E_2^{N}(S,T)&=&1-\tfrac{1}{(2MN)^2}\sum_{n,n'}|{\rm Tr}[U^{n-n'}]|^2\label{E2a}\\
&=&{\textstyle 1-\frac{1}{N}[1+2\!\!\!\sum\limits_{n=1}^{\lfloor \frac{N-1}{2}\rfloor}\!\!(1-\frac{2n}{N}) P^2_n(\cos^2\theta)].}\;\;\;\;\;\;
\label{E2Nn}
\end{eqnarray}
\end{subequations}
As a check, it is verified  that for $\theta=0$, we 
obtain from \eqref{E2Nn} maximum entropy $\forall$ $N$ ($P_n(1)=0$ for $n\geq 1$): 
\begin{equation}
E_2(S,T)=1-\frac{1}{N},\;\;\;\;(\theta=0)\,,
\end{equation}
in agreement with previous considerations (any entropy is obviously also maximum in this case). And for $\theta=\pi/2$,  we recover from \eqref{E2Nn} Eq.\ \eqref{E2th2}  for both $N$ even or odd, as $P_n(0)=1$ $\forall$ $n\geq 1$. Exact results as a function of $N$ and $\theta$ are depicted in Figs.\ \ref{fig2}--\ref{fig3}.

For large $N$, we may use Eq.\ \eqref{N} for $n\neq 0$ and approximate  the sum over $n$ in \eqref{E2Nn} by an integral over $u=2n/N$,  which leads to the asymptotic expression 
\begin{eqnarray}
\!\!\!\!\!E_2^{N}(S,T)&\approx&1-{\textstyle\frac{1}{N}\left[1+\frac{2}{\pi}\left(\ln \frac{N}{2}+{\rm Si}(4\theta)-c\right)\tan\theta\right]},\;\;\;\; 
\end{eqnarray}
where ${\rm Si}(x)=\int_0^x\frac{\sin t}{t}dt$ and  $c=1+\frac{\pi}{2}$ (neglecting terms $O(N^{-2})$). The deviation from  maximum entropy $S_2^{\rm max}=1-\frac{1}{N}$ is then $O(\frac{\ln N}{N})$ and proportional to $\tan\theta$, in agreement with previous considerations. 
This result can  also be obtained from
\eqref{S2N} using $\frac{\sin^2(Nu/2)}{N\sin^2 u/2}{\mathop{\rightarrow}\limits_{N\rightarrow\infty}}2\pi\delta(u)$ for $|u|<\pi$ and integrating over $\phi=\frac{2\pi k}{m}$. An exact summation using \eqref{N} is given in the appendix. 
\subsubsection{Entanglement spectrum}
For an initially localized particle, we may also examine the entanglement spectrum, i.e., the common eigenvalues of the reduced densities \eqref{rhor} which  determine the entropies of Figs.\ \ref{fig2}--\ref{fig3},  by diagonalizing the overlap matrix  (Eqs.\ \eqref{cw}--\eqref{pl})
\begin{equation}
\langle\Psi_{n'}|\Psi_{n}\rangle=\left\{
\begin{array}{ccl}P_{\frac{|n-n'|}{2}}(\cos^2\theta)&,&n-n'\;{\rm even}\\ 0&,&n-n'\;{\rm odd}\end{array}\right.\,, \label{ovv}\end{equation}
for $0\leq n,n'\leq N-1$. This leads to two similar blocks 
($n,n'$ even or odd respectively), identical for $N$ even. 
\begin{figure}[h]
    \centering
    \hspace*{-0.5cm}\includegraphics[width=.5\textwidth]{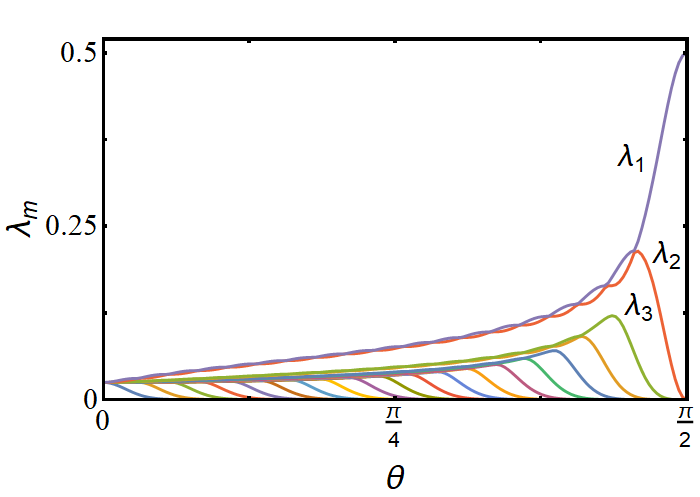}
    \caption{Entanglement spectrum  of the history 
 state (eigenvalues $\lambda_1\geq\lambda_2\geq\ldots$ of the reduced densities $\rho_S$ or $\rho_T$) obtained from the overlap matrix \eqref{ovv} for an initially localized particle and $N=40$  as a function of the coin operator parameter $\theta$.}
    \label{fig4}
\end{figure}

For sufficiently large $N$  the eigenvalues of each block (identical for $N$ even) come essentially in almost degenerate pairs for $\theta$ not close to $\pi/2$,  with the number of non-negligible eigenvalues  decreasing 
with increasing $\theta$, as depicted in Fig.\ \ref{fig4}. The largest eigenvalue lies  close to the Gershgorin upper bound, i.e., using Eq.\ \eqref{N}, 
\begin{subequations}
\label{ovs}
\begin{eqnarray}\lambda_{\rm max}(\theta)&\approx &{\textstyle\frac{1}{N}[1+\sqrt\frac{\tan\theta}{\pi}
\sum\limits_{n=1}^{\frac{N}{2}-1}\frac{|\cos(2n\theta+\pi/4)|}{\sqrt{n}}]}\\&\approx&{\textstyle\frac{1}{N} [1+\sqrt{\frac{\tan\theta}{2\pi}}H_{\frac{1}{2}}(\tfrac{N}{2}-1)]}\\
&\approx&{\frac{1}{N}+
\sqrt{\frac{\tan\theta}{N\pi}}\left[1+
\frac{\zeta(\frac{1}{2})}{\sqrt{2N}}\right]}\,,\end{eqnarray}
\end{subequations}
where $H_{1/2}(m)=\sum_{n=1}^m\frac{1}{\sqrt{n}}$ ($\approx 2\sqrt{m}+\zeta(\frac{1}{2})$ for large $m$) is the generalized harmonic number ($\zeta(1/2)\approx -1.46$ is the Riemann zeta function at $1/2$).  Thus, there is always a  deviation 
$O(N^{-1/2})$ from the maximally mixed distribution, proportional to $\tan^{1/2}\theta$.

\section{Relation with operator entanglement \label{IV}} 
\subsection{Entanglement of unitary operators}
We will show here that the system-time entanglement entropy for the initially localized particle, which is independent of the initial (real) spin state, 
is  the entanglement entropy of the global unitary  operator which generates the quantum walk.

First, let us consider a complete set  
 of local orthogonal operators $O_i^S$  ($O_j^T$)  of the system (clock)    satisfying 
 \begin{equation} {\rm Tr}_{S}\,[O_i^{S\dag} O_{i'}^S]=d_S\delta_{ii'}\,,\;\;\;\; {\rm Tr}_{T}\,[O_j^{T\dag} O_{j'}^T]=d_T\delta_{jj'}\,,\label{ortg}\end{equation} 
 where $d_{S(T)}$ is the Hilbert space dimension of $S$ ($T$). Any operator ${\cal U}$ on the whole system can  be expanded as  
\begin{equation}
{\cal U}=\sum_{i,j}{M}_{ij}\,O_i^S\otimes O_j^T\,,\label{Uxp}
\end{equation}
where  ${M}_{ij}=\frac{1}{d_S d_T}{\rm Tr}\,[(O_i^\dag\otimes O_j^\dag)\,{\cal U}]$. Then  
$\frac{1}{d_S d_T}{\rm Tr}\,[{\cal U}^\dag{\cal U}]=\sum_{i,j}|{M}_{ij}|^2={\rm Tr}[{M}^\dag {M}]$. If ${\cal U}$ is unitary, ${\rm Tr}\,[{M}^\dag {M}]=1$ and  the   $|{M}_{ij}|^2$ become  standard probabilities.   

Hence, in the same way as done for the history state, through the SVD   ${M}=VDW^\dag$, with $D_{mm'}={\lambda}_m\delta_{mm'}$ and $V$, $W$ unitary,  we can also obtain the Schmidt representation  of the operator ${\cal U}$, 
\begin{equation}
{\cal U}=\sum_{m=1}^{\tilde{n}_s}\sqrt{\tilde\lambda_m}\, \tilde{O}_m^S\otimes \tilde{O}_m^T\,,\label{SVD2}
\end{equation}
where $\lambda_m$ are the eigenvalues of   $ M^\dag M$ or $ M M^\dag$ and  
$\tilde{O}_m^{S}=\sum_i V_{im}O_i^{S}$, 
$\tilde{O}_m^{T}=\sum_j W^*_{jm}O_j^{T}$ 
 are new local orthogonal operators satisfying \eqref{ortg}, 
  with $\tilde{n}_s$ the rank of $\tilde M$.  If ${\cal U}$ is unitary the eigenvalues $\lambda_m$ are again  standard probabilities ($\lambda_m\geq 0$, $\sum_m \lambda_m=1$).  Thus, for a general trace-form entropy, the entanglement entropy of the unitary operator ${\cal U}$ can  be defined as \begin{equation}
E({\cal U})=\sum_{m}f(\lambda_m)\,,\label{OE}
\end{equation}
vanishing iff ${\cal U}$ is a product of local unitaries ($\tilde{n}_s=1$). 

The analogy with state entanglement is straightforward if the Choi isomorphism  for representing operators is employed. 
Any operator ${\cal U}$ on  ${\cal H}_{ST}={\cal H}_S\otimes {\cal H}_T$  can be associated to a pure state $|{\cal U}\rangle$ in ${\cal H}_{ST}\otimes {\cal H}_{ST}$ given by 
\begin{equation}
|{\cal U}\rangle=({\cal U}\otimes\mathbbm{1}_{ST})|\mathbbm{1}\rangle=\frac{1}{\sqrt{d_S d_T}}\sum_{\mu}({\cal U}|\mu\rangle)\otimes |\mu\rangle\label{STU}
\end{equation}
where $|\mathbbm{1}\rangle=\frac{1}{\sqrt{d_S d_T}}\sum_\mu |\mu\rangle\otimes|\mu\rangle$ is a maximally entangled state in ${\cal H}_{ST}\otimes {\cal H}_{ST}$. In this way, an exact map for inner products is obtained: 
\begin{equation}
\langle {\cal U}'|{\cal U}\rangle=\frac{1}{d_S d_T}{\rm Tr}\,[{\cal U}^{\prime\dag} {\cal U}]\,,\end{equation}
with $\langle {\cal U}|{\cal U}\rangle=1$ for a unitary operator ${\cal U}$. 
Thus, Eqs.\ \eqref{Uxp}--\eqref{SVD2} can be recast, noting that $|\mathbbm{1}\rangle=|\mathbbm{1}_S\rangle\otimes|\mathbbm{1}_T\rangle$,  as 
\begin{eqnarray}
|{\cal U}\rangle&=&\sum_{i,j} M_{ij}|O_i^S\rangle\otimes|O^T_j\rangle 
=\sum_m \sqrt{\lambda_m}|\tilde{O}_m^S\rangle\otimes|\tilde{O}^T_m\rangle\;\;\;\;\end{eqnarray}
with $|O_i^{S}\rangle=(O_i^{S}\otimes \mathbbm{1}_S)|\mathbbm{1}_S\rangle=\frac{1}{d_S}\sum_\mu (O_i^S|\mu_S\rangle)\otimes |\mu_S\rangle$  and similarly for $|O_j^T\rangle$, 
such that $ M_{ij}=(\langle O_i^S|\otimes \langle O_j^T|)|{\cal U}\rangle$. 
We can now rewrite \eqref{OE} as 
\begin{equation}
E({\cal U})=S(\rho^{\cal U}_S)=S(\rho^{\cal U}_T)\,,
\end{equation}
where $\rho^{\cal U}_{S(T)}={\rm Tr}_{T(S)}\, |{\cal U}\rangle\langle {\cal U}|=\sum_{m}\lambda_m|\tilde{O}_m^{S(T)}\rangle\langle \tilde{O}_m^{S(T)}|$   are the local reduced densities derived from $|{\cal U}\rangle$.
In particular,  the quadratic entropy becomes 
\begin{align}E_2({\cal U})&=1-{\rm Tr}\,(\rho_{S(T)}^{\cal U})^2=1\!-\!{\rm Tr}\,[( M M^\dag)^2]\,.\end{align}

\subsection{Operator entanglement and quantum walk}
If a system initially in a state $|\Psi_0\rangle$ undergoes a discrete unitary evolution through times $n=0,\ldots,N-1$ and states $|\Psi_n\rangle= U_n|\Psi_0\rangle$, 
its history state 
can be generated from an initial system-clock product state  as 
(see Fig.\ \ref{fig1}) 
\begin{equation}|\ppsi\rangle=\frac{1}{\sqrt{N}}\sum_{n=0}^{N-1}U_n|\Psi_0\rangle\otimes |n\rangle 
={\cal W}(|\Psi_0\rangle\otimes |\tilde{0}\rangle)\,,\label{HS2}\end{equation}
where $|\tilde{0}\rangle=H^{\otimes m}|0\rangle=\frac{1}{\sqrt{N}}
\sum_{n=0}^{N-1} |n\rangle$ and 
\begin{equation}{\cal W}=\sum_{n=0}^{N-1} U_n\otimes |n\rangle\langle n|\label{Wo}
\end{equation}
is a controlled-$U_n$ unitary operator on the whole system. Its state representation \eqref{STU} is 
\begin{equation}|{\cal W}\rangle=\frac{1}{\sqrt{N}}\sum_{n=0}^{N-1}|U_n\rangle\otimes|T_n\rangle\label{Ws}\end{equation}
where $|T_n\rangle=(\sqrt{N}|n\rangle\langle n|\otimes \mathbbm{1})|\mathbbm{1}\rangle=|n\rangle\otimes |n\rangle$  and $|U_n\rangle=\frac{1}{\sqrt{d_S}}\sum_\mu (U_n|\mu_S\rangle)\otimes |\mu_S\rangle$.  Therefore, 
the unitary operator ${\cal W}$ generating the history state from a product state  can itself be represented as an operator history state \eqref{Ws}. 

 The reduced operator state of the clock is  here 
\begin{equation}
\rho^{\cal W}_T={\rm Tr}_{S}|{\cal W}\rangle\langle {\cal W}|=\frac{1}{N}\sum_{n,n'}\langle U_n|U_{n'}\rangle |T_n\rangle\langle T_n|\,,
\end{equation}
 in  full analogy with \eqref{rhoT}, showing again that the entanglement of the generating operator \eqref{Wo}--\eqref{Ws},  
 \begin{equation} E({\cal W})=S(\rho_S^{\cal W})=S(\rho_T^{\cal W})\,,
 \label{Ew3}\end{equation}
is fully determined by the overlaps $\langle U_{n'}|U_n\rangle$.  
In the present random walk, $U_n=U^n$ and 
\begin{equation}\langle U_{n'}|U_{n}\rangle=\frac{1}{2M}{\rm Tr} [U^{n-n'}]\,,\end{equation}
is exactly the overlap   \eqref{cw}-\eqref{pl}--\eqref{ovv} between the evolved system states for an initially localized particle with real initial spin state.  

Thus, the operator entanglement \eqref{Ew3}  {\it is exactly that of the previous system-clock history state for the initially localized particle}, for {\it any} choice of entropy.  In particular, the quadratic operator entanglement 
 \begin{equation}
 E_2({\cal W})=1-{\rm Tr}\,(\rho_T^{\cal W})^2=1-\frac{1}{N^2}\sum_{n,n'}|\langle U_{n'}|U_n\rangle|^2\,,  \label{SW}\end{equation}
 is just the quadratic $S$--$T$ entropy \eqref{E2N}.   And the entanglement spectrum of ${\cal W}$ coincides with that of the history state  (Fig.\ \ref{fig4}). 
 
The quadratic entropy \eqref{SW} has an additional meaning: It  determines the {\it entangling power} of the unitary operator ${\cal W}$, i.e.\ the average  quadratic entanglement $E_2(S,T)$ it generates when applied to initial product system-clock states: 
\begin{eqnarray}
\langle E_2(S,T)\rangle&=&\int_{\cal H} (1-{\rm Tr}\,\rho_{S}^2)\,d\Psi_0 
\nonumber\\&=&\frac{d_S}{d_S+1} E_2({\cal W}),\label{EP}
\end{eqnarray}
with the average taken over the whole set of  initial system states $|\Psi_0\rangle$ with the Haar measure $d\Psi_0$ \cite{BR.18}. Thus, for $d_S\gg 1$ the  quadratic $S$--$T$ entanglement entropy  \eqref{E2N}, 
equal to \eqref{SW}, is very close to the average value \eqref{EP}. 

\subsection{Spin history states in the quantum walk}

If we now consider as system 
only the spin degree of freedom, we first notice from  Eq.\ \eqref{Un} that for a  particle initially localized at $x=0$,  the  state after $n$ steps \begin{equation}|\Psi_n\rangle =\frac{1}{\sqrt{M}}\sum_k |k\rangle\otimes U_k^n|\chi_0\rangle\,,\label{psis}\end{equation}  
is  exactly a {\it spin history state}, with respect to the momentum states $|k\rangle$. 
The ``evolution'' operators  are  here the $k$-projected unitaries $U_k^n$ acting on the spin, 
having a nontrivial $k$-dependence. Thus, the spin-position entanglement at step $N$ is that of a spin history state. The same holds for any initial localization $x_0$ ($|k\rangle\rightarrow  e^{-i2\pi kx_0/M}|k\rangle$). 

Therefore, its average $\langle E_2(s,p)\rangle$ over all initial spin states is determined by the entanglement of the operator $U^n$ generating such spin history [from the initial product state $|\Psi_0\rangle=(\frac{1}{\sqrt{M}}\sum_{k}|k\rangle)|\chi_0\rangle$], 
itself an operator history state, as is apparent  from Eq.\ \eqref{Ukk}: 
\begin{equation}
U^n = \sum_{k=0}^{M-1} |k\rangle\langle k|\otimes U_k^n
\label{UU}\,.\end{equation}
The associated operator state is  $\frac{1}{\sqrt{M}}\sum_k|P_k\rangle|U_k^n\rangle$ with $|P_k\rangle=|k\rangle\otimes|k\rangle$ and $|U_k^n\rangle=\frac{1}{\sqrt{2}}\sum_{\nu=\pm}U_k^n|\nu\rangle|\nu\rangle$ ($|\nu=\pm\rangle$ are orthogonal spin states).  Its entanglement is then determined by the overlaps (see Eqs.\ \eqref{10} and   \eqref{uknx} below) 
\begin{eqnarray} \langle U_{k'}^n|U_k^n\rangle&=&\tfrac{1}{2}{\rm Tr}\,[U_{k'}^{-n}\,U_{k}^n]\,.
\end{eqnarray}

\begin{figure}[t]
    %\centering
   \hspace*{-.8cm} \includegraphics[width=.5\textwidth]{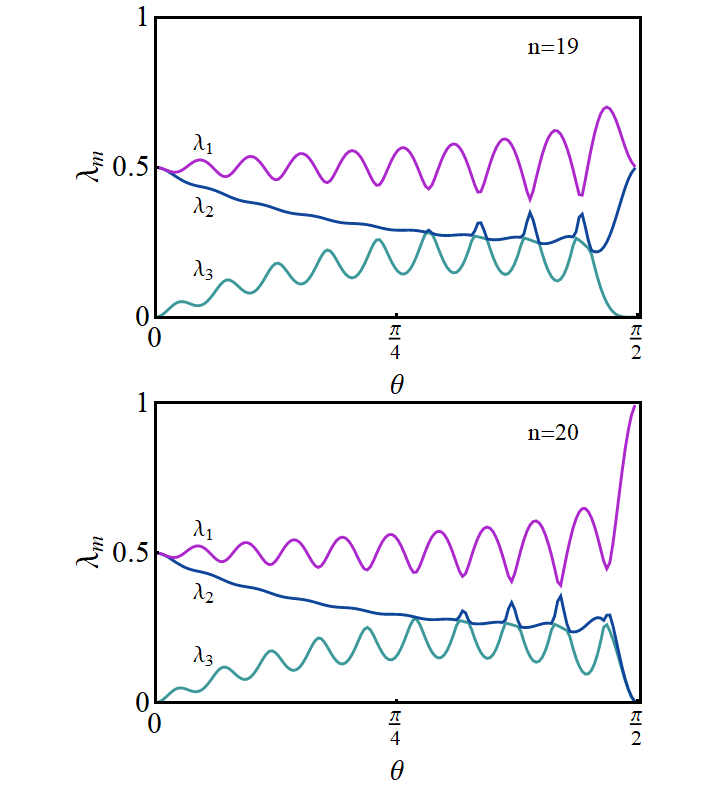}
    \caption{Entanglement spectrum of the unitary operator (\ref{UU}) for an  odd (top) and even (bottom) $n$ as a function of the coin operator parameter $\theta$.}
    \label{fig5}
\end{figure}

While the unitaries  $U_k^n$ belong to a four-dimensional space spanned by the identity and the three Pauli operators,  limiting the Schmidt rank of \eqref{UU} to $n_s\leq 4$  $\forall$ $M$,   they actually  live  within a {\it three-dimensional} subspace 
spanned by the identity $\mathbbm{1}_s$, the coin operator $\sigma_\theta\equiv C$ and the orthogonal Pauli operator $\sigma_y$: From  \eqref{5}--\eqref{8} we   see that 
\begin{equation}
U_k=\cos\phi_k\,\sigma_\theta-i\sin\phi_k \exp[i\theta\sigma_y]\,,
\label{Uspan}\end{equation}
since $\sigma_z\sigma_\theta=\cos\theta\,\mathbbm{1}_s+i\sin\theta\,\sigma_y$. And from \eqref{10b} and \eqref{Uspan} it follows  that 
\begin{subequations}\label{69}
\begin{eqnarray}|s_k^\pm\rangle\langle s_k^\pm|
&=&\tfrac{1}{2}(\mathbbm{1}\pm\sigma_k)\,,\\
\sigma_k&=&\frac{\cos\phi_k}{\cos\omega_k}\sigma_\theta+\frac{\sin\phi_k\sin\theta}{\cos\omega_k}\sigma_y\,,\end{eqnarray}
\end{subequations}
 with $\sigma_k^2=\mathbbm{1}$. Hence all powers 
\begin{subequations}\label{uknx}\begin{eqnarray}\label{eqnk} U_k^n&=&e^{-in\omega_k}|s_k^+\rangle\langle s_k^+|+(-1)^ne^{in\omega_k}|s_k^-\rangle\langle s_k^-|\nonumber\\
&=&\cos(n\omega_k)\mathbbm{1}-i\sin (n\omega_k)\sigma_k\;\;\;\;\;\;\;(n\;{\rm even})\,,\label{uknxe}\\
&=&-i\sin(n\omega_k)\mathbbm{1}+\cos(n\omega_k)\sigma_k\;\;\;\;\;(n\;{\rm odd})\,,\label{uknxo}\end{eqnarray}
\end{subequations}
are spanned just by $\{\mathbbm{1}_s,\sigma_\theta,\sigma_y\}$ $\forall$ $k,n$, with ${\rm Tr}\,\sigma_\theta\sigma_y=0$.  This entails a Schmidt rank $n_s\leq 3$ of the $n^{\rm th}$ power  \eqref{UU}, implying at most  $3$ non-zero eigenvalues ${\lambda}_m$ in its associated entanglement spectrum, 
as seen in Fig.\ \ref{fig5}, 
and a von Neumann entanglement entropy $E(U^n)\leq \log_2 3$.

For $U$ itself ($n=1$) just $\{\sigma_\theta,e^{i\theta\sigma_y}\}$ are needed as seen from \eqref{Uspan}, implying $n_s=2$, as also evident from the original expression  \eqref{Uy}. Its entanglement spectrum is just $(1/2,1/2,0)$. However, for $n\geq 2$ the rank is $n_s=3$  if  $\theta\in(0,\pi/2)$ (and $M>3$). Exceptions occur for $\theta=0$, in which case $\omega_k=\phi_k$ and $\sigma_k=\sigma_\theta=\sigma_z$ $\forall\,  k$, leading to $n_s=2$ and an entanglement spectrum $(1/2,1/2,0)$ $\forall$ $n\geq 1$, 
as verified in Fig.\ \ref{fig5}, and also for $\theta=\pi/2$, where $\omega_k=0$ $\forall$ $k$ and $\sigma_\theta=\sigma_x$, leading to $n_s=1$ $\forall$ $n$ even (as $U_k^n=\mathbbm{1}$ $\forall$ $k$) and $n_s=2$ for $n$ odd (as $U_k^n=\sigma_k$), with spectrum $(1/2,1/2,0)$, as also seen in Fig.\ \ref{fig5}.  

\begin{figure}[t]
    \centering
    \includegraphics[width=.45\textwidth]{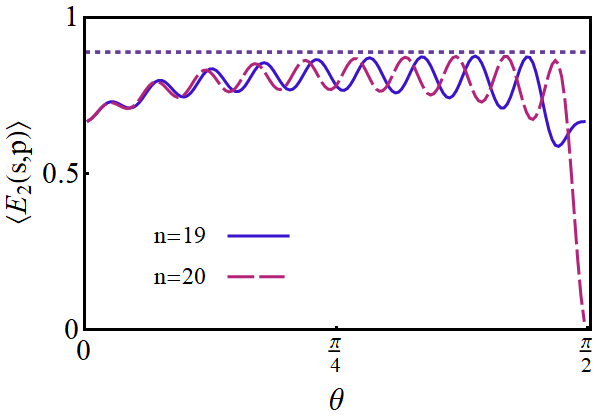}
    \caption{The average over all initial spin states of the   spin-position  quadratic entanglement entropy  $\langle E_2(s,p)\rangle=(2/3)E_2(U^n)$ (Eq.\ \eqref{Unep}), as a function of the coin operator parameter $\theta$, for the even and odd cases of Fig.\ \ref{fig5}. The dotted line indicates the upper bound $8/9$.}
    \label{fig6}
\end{figure}

Therefore, the entangling power of $U^n$ remains bounded by this rank $n_s\leq 3$:  Rescaling the quadratic entropy as $S_2(\rho)=2(1-{\rm Tr}\,\rho^2)$, such that  $S_2(\rho)=1$ for a maximally mixed single spin state, it  implies  $E_2(U^n)\leq 4/3$. 
Hence, applying the relation  \eqref{EP} to the spin history, 
  the average over all initial spin states of the spin-position entanglement after $n$ steps satisfies  
\begin{equation}\langle E_2(s,p)\rangle=(2/3)E_2(U^n)\leq 8/9\,.\label{Unep}\end{equation}  
  This bound is obviously lower than the maximum   $\langle E_2(s,p)\rangle=1$ reached for a full rank maximally entangled  operator $U_n$  ($E_2(U_n)=3/2$ for the rescaled $S_2$), for  which  $|\Psi_n\rangle$ in \ \eqref{psis}  would be {\it maximally entangled}  ($E_2(s,p)=1$)  for {\it any} spin state $|\chi_0\rangle$,  in agreement with  the general results of \cite{BR.18}. The variation of  $\langle E_2(s,p)\rangle\propto E_2(U)$  
  with $\theta$ is depicted in Fig.\ \ref{fig6}. 
  
Similarly, the  full  history state \eqref{psih} for the initially localized particle can also be regarded, when viewed from the spin,  as a spin history state with respect to a {\it composite clock},   comprising both the original clock  and the position degrees of freedom:
\begin{equation}
|\ppsi\rangle=\frac{1}{\sqrt{NM}}\sum_{k,n}|kn\rangle\otimes  U_{k}^n|\chi_0\rangle,\label{psih2}
\end{equation}
where $|kn\rangle=|k\rangle\otimes|n\rangle$  and for clarity we have altered the order in the tensor product. In this case the unitary operator generating the full spin history is 
\begin{equation}
{\cal W}_s=\sum_{k,n}|kn\rangle\langle kn|\otimes U_k^n\,.\label{WW}
\end{equation}
The average (over all initial spin states) $\langle E_2(s,pT)\rangle$ of the spin--rest entanglement in the state \eqref{psih2}  is then determined by the entanglement of the operator \eqref{WW}, 
in turn determined by the full set of overlaps 
\begin{eqnarray}
\langle U_{k'}^{n'}|U_k^n\rangle&=&\tfrac{1}{2}{\rm Tr}\,[U_{k'}^{-n'}\,U_{k}^n]\,,
\end{eqnarray}
for $0\leq k,k'\leq M$, $0\leq n,n'\leq N-1$.

\begin{figure}[t]
    \centering
    \includegraphics[width=.45\textwidth]{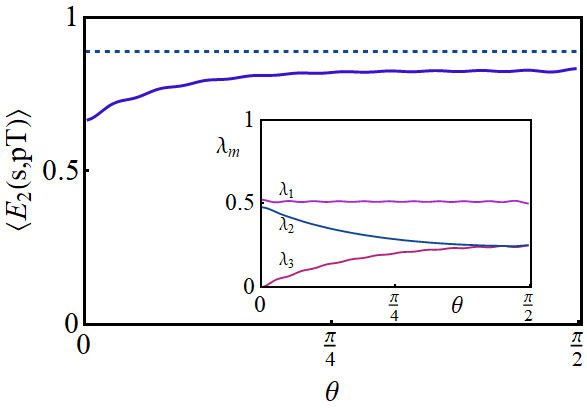}
    \caption{Average over all  initial spin states of the  spin--rest quadratic  entanglement entropy, $\langle E_2(s,pT)\rangle=(2/3)E_2({\cal W}_s)$ (Eq.\ \eqref{Wsep}), as a function of the coin  operator parameter $\theta$, for $N=20$. The dotted line indicates the upper bound $8/9$.  Inset: Entanglement spectrum of the unitary operator ${\cal W}_s$, Eq.\  (\ref{WW}) generating the full spin history, as a function of the coin  operator parameter $\theta$.}
    \label{fig7}
\end{figure}

Nonetheless, since all $U_k^n$ are spanned just by  $\{\mathbbm{1}_s,\sigma_\theta,\sigma_y\}$ (Eq.\ \eqref{Uspan}),  the Schmidt rank of \eqref{WW} is again $n_s=3$ (or $n_s\leq 3$ in general), and  the previous bound, i.e.\   
$E_2({\cal W}_s)\leq 4/3$ 
still holds. This entails again 
\begin{equation}\langle E_2(s,pT)\rangle=(2/3)E_2({\cal W}_s)\leq 8/9\,.\label{Wsep}\end{equation}
This average is depicted as a function of $\theta$ in Fig.\ \ref{fig7}, together with the entanglement spectrum of ${\cal W}_s$. 

\section{Conclusions\label{V}}
We have analyzed the history state formalism in the context of discrete QW. The history state captures the whole evolution of the system, enabling for instance the evaluation of time averages as single quantum expectation values. It satisfies a timeless eigenvalue equation and can be generated through a quantum circuit from an initial system-clock product state. The associated system-clock entanglement entropy  is a measure of the number of orthogonal system  states visited in the whole QW and is fully determined by the overlaps between the evolved states. Stationary system states then lead to  {\it separable} history states, while QW in which the system evolves into a new orthogonal state at each step correspond to maximally entangled history states. 

We have then shown that in  one dimensional QW with real Hadamard-type coin operators,  such entanglement is strictly independent of the initial spin orientation for real initial states with definite position parity. We also analyzed its connection with operator entanglement, showing that 
in the standard case of an initially localized particle it coincides exactly with the entanglement of the unitary operator generating the whole QW. Exact analytic results for overlaps and quadratic entropies as a function of the number $N$ of steps and the coin parameter were as well derived,  including asymptotic expressions for large $N$. The latter  show  a monotonously decreasing entropy with increasing coin  operator parameter  $\theta\in[0,\pi/2]$, with $S_2(0)-S_2(\theta)\propto\tan\theta\frac{\ln N}{N}$ for the quadratic entropy. 
 The associated history state entanglement spectrum shows,  accordingly, a decreasing rank for increasing $\theta$, with a deviation $O(\frac{\tan\theta}{N})^{\frac{1}{2}}$ from $N^{-1}$ in the largest eigenvalue.  

Finally, we have examined the QW from the spin perspective, showing that for an initially localized particle it is also  described by a history state with a composite clock and a  $k$-dependent unitary. The average over all initial spin states of the ensuing spin-rest entanglement 
can then be related to that of the global unitary generating this history. For the present Hadamard-type coin operator it 
 has a limited rank with just three non-zero eigenvalues in its entanglement spectrum, leading to a bounded average spin-rest entanglement. 

 In summary,  the history state formalism 
 provides a novel perspective for analyzing QW. 
 The associated system-time entanglement entropy constitutes a new measure for characterizing the whole evolution, which could be employed, for instance, in relation with the identification of dynamical phase transitions \cite{Wang.19} and topological phases \cite{Kita.10,Xu.18,car.20,QQW.20,col.22}, of great current interest.
 The formalism also opens the way to efficient direct evaluation of time averages and quadratic entanglement entropies through its simulation in a quantum circuit \cite{BRGC.16,BR.18,DP.19}, while a variational determination of the history state becomes as well possible.  Finally, the extension of the formalism to open systems with nonunitary dynamics and  to more complex scenarios is in principle feasible and is currently under investigation. 
 
\acknowledgments 
Authors acknowledge support from CONICET (F.L., N.C.\  and A.P.B.) and CIC (R.R.) of Argentina.  Work supported by
CONICET PIP Grant No. 11220200101877CO.

\appendix

\section{Quadratic entanglement entropy summation}
The sum for the quadratic  entanglement entropy \eqref{E2Nn} can be done exactly for large $n$ using  the approximation  
\eqref{N} and neglecting terms  $O(n^{-\frac{3}{2}})$.  This yields 
\begin{eqnarray}
E_2(S,T)&\approx& 1-[1+\tfrac{\tan\theta}{\pi}F(N,\theta)]/N\,,\end{eqnarray}
with 
\begin{eqnarray}
F(N,\theta)&=&2\sum_{n=1}^\eta(\tfrac{1}{n}-\tfrac{2}{N})\cos^2(2n\theta+\pi/4) \\
&=&H(\eta)+{\rm Im}[e^{4i\theta\eta}L(e^{4i\theta},\eta)]+2\theta-\tfrac{\pi}{2}\nonumber\\
&&-\tfrac{2}{N}[\eta-\tfrac{\sin(2\theta(\eta+1))\sin(2\theta\eta)}{\sin(2\theta)}]\,,
\end{eqnarray}
where  $\eta=\lfloor\frac{N-1}{2}\rfloor$,  $H(\eta)=\sum_{n=1}^\eta 1/n$ is the harmonic number and $L(z,\eta):=\sum_{k=1}^\infty{z^k}/(k+\eta)$ is  directly related to the Lerch zeta function \cite{Abra.65}. 

%\bibliography{biblio}

\end{document}